\documentclass[12pt,preprint]{aastex}  

\def\degpnt{^{\circ}\kern-1.7mm.\kern+.35mm}
\def\arcpnt{"\kern-1.7mm.\kern+.35mm}
\def\minpnt{'\kern-1.0mm.\kern+.30mm}

\def\lessim{\mathrel{\hbox{\rlap{\hbox{\lower4pt\hbox{$\sim$}}}\hbox{$<$}}}}
\def\grtsim{\mathrel{\hbox{\rlap{\hbox{\lower4pt\hbox{$\sim$}}}\hbox{$>$}}}}

\shorttitle{Novae in M101}
\shortauthors{Coelho, Shafter, \& Misselt}

\begin{document}


\title{The Rate and Spatial Distribution of Novae in M101 (NGC~5457)}


\author{E.~A. Coelho and A.~W. Shafter}
\affil{Department of Astronomy and Mount Laguna Observatory, San Diego State University, San Diego, CA 92182}
\email{ecoelho@sciences.sdsu.edu, aws@nova.sdsu.edu}
\and
\author{K.~A. Misselt}
\affil{Steward Observatory, University of Arizona, 933 North Cherry Avenue, Tucson, AZ 85721}
\email{misselt@as.arizona.edu}



\begin{abstract}

A new multi-epoch H$\alpha$ imaging study of M101 (NGC~5457) has been
carried out as part of a larger campaign
to study the rate and stellar population of extragalactic novae.
The survey yielded a total of 13 nova detections from 10 epochs
of M101 observations
spanning a three year period. After correcting for the temporal
coverage and survey completeness,
a global nova rate of $11.7^{+1.9}_{-1.5}$ yr$^{-1}$ is found.
This value corresponds to a luminosity-specific
nova rate of $1.23\pm0.27$ novae per year per $10^{10}~L_{\odot,K}$
when the $K$ luminosity is derived from
the $B-K$ color, or $1.94\pm0.42$ novae per year per $10^{10}~L_{\odot,K}$
when the $K$ magnitude from the Two Micron All Sky Survey is used.
These values are consistent with previous estimates by Shafter et al.
that were based on more limited data.
The spatial distribution of the combined nova sample
from the present survey and from the earlier Shafter et al. survey
shows that the specific frequency of novae closely follows the integrated
background light of the galaxy.

\end{abstract}




\section{Introduction}

Novae occur in semi-detached binary systems in which a white dwarf (WD) primary
accretes material from a Roche-lobe filling late-type companion (see
Warner 1995 for a review). For sufficiently low mass accretion rates,
the accreted matter builds up on the surface of the WD 
under degenerate conditions.
When the Fermi temperature is reached at the base of
the accreted layer, a thermonuclear runaway (TNR) ensues,
blowing off the outer layers of the star, and 
resulting in an outburst of 10--20 magnitudes.
In the resulting explosion, novae reach peak luminosities
of up to $M_{V} \sim -10$, making them visible in external galaxies 
up to and beyond the Virgo cluster.
The high luminosity, coupled with a well-known relationship between
a nova's maximum magnitude and its rate of decline, the MMRD relation
(Zwicky 1936, Mclaughlin 1945, Downes \& Duerbeck 2000),
has enabled
novae to be useful as an extragalactic distance indicator
(e.g. Pritchet \& van den Bergh 1987, Della Valle \& Gilmozzi 2002).
However, difficulties in calibrating the MMRD relationship,
as well as uncertainties about
the uniformity of the progenitor populations in different galaxy types have 
hindered the use of novae
as reliable distance indicators (Ferrarese et al. 2003).
Recent studies have focused mainly on the 
value of using novae to probe extragalactic stellar populations.

Galactic nova studies have suggested that there are actually two
distinct classes of classical
novae, a disk population and a bulge population (e.g., Duerbeck 1990,
Della Valle et al. 1992; Della Valle \& Livio 1998).
As shown by Della Valle et al. (1992),
``fast" novae\footnote{Fast novae can be loosley defined as those
novae which decline at least two magnitudes within 12 days
(i.e. $t_2\leq12$~days).} tend to be associated with a disk
population, having an average scale height above the Galactic plane
generally less that that of slower novae.
At about the same time Williams (1992) pointed out that novae could also
be segregated into two classes based upon their spectroscopic
characteristics. Novae whose spectra display
prominent Fe~II lines (the Fe~II novae)
tend to evolve more slowly, have lower
expansion velocities, and have a lower level of ionization
compared with novae that exhibit strong lines of He and N (the He/N novae).
In addition, these He/N novae often display very strong neon lines,
which suggests that the seat of the eruption is a relatively
massive ONe white dwarf. Such novae do not appear to produce the
copious carbon-rich dust that is often formed in nova ejecta arising
from the lower mass CO white dwarfs (Gehrz et al. 1998).
Additional support for the two-population scenario was
provided by Della Valle \& Livio (1998), who looked into the
spatial distribution of a sample of 27
Galactic novae with known spectral
class and reliable distance estimates.
They noted that Galactic novae that could be
classified as He/N were more concentrated to the Galactic plane,
and tended to
be faster and more luminous compared with their Fe~II counterparts.

Despite the progress made in understanding nova populations
with Galactic data, observations of novae in nearby galaxies offer
several advantages over Galactic studies.
For a given galaxy, the nova sample is essentially equidistant,
with constant (and usually minimal) foreground reddening,
making it easier to study
relative nova luminosities and fade rates. For spatially-resolved
spiral systems it is often possible to segregate novae from
different spatial positions within a galaxy (e.g. between
the bulges and disks of spirals), revealing possible differences
between novae from differing stellar populations. Perhaps
most importantly, extragalactic nova observations make it
possible to study novae from isolated stellar populations directly
by comparing their properties in galaxies along the Hubble sequence
from ellipticals to irregulars.

A number of studies have looked at the question of whether 
the luminosity-specific nova rate (LSNR)
of a galaxy varies with Hubble type. In these studies, the
nova rate of a galaxy is normalized to its infrared luminosity
(usually $K$-band), which acts as a proxy
for stellar mass.
In the first such study, Ciardullo et al. (1990b) compiled a list of
galaxies with measured nova rates and concluded
that the LSNR was essentially independent of Hubble type. Shortly
thereafter, Della Valle et al. (1994) re-analyzed much of the same data
(using different estimates for completeness and for
the normalization of the nova rate by galaxy luminosity) and came
to a different conclusion. Namely, that the LSNRs of
the late-type, low-mass galaxies in the local group (SMC, LMC, and M33)
were significantly higher than those of the other, earlier-type galaxies,
such as the Virgo ellipticals.

In an important paper, Yungelson et al. (1997) published the results of
population synthesis models demonstrating that the
LSNR
in a galaxy should be strongly dependent on the galaxy's star
formation history. There are two reasons for this dependence.
The first is that the mean mass of the accreting
white dwarfs in newly-formed nova binaries declines with increasing
time elapsed since the formation of the zero-age main sequence system
(Tutukov \& Yungelson 1995). Since the frequency of nova eruptions
increases with increasing white dwarf mass (Ritter et al. 1991; Livio 1992),
galaxies experiencing recent ($1-2\times10^9$~yr)
star formation, such as late-type spirals, should be
more prolific nova producers compared with galaxies that
formed the bulk of their stars in a burst of star formation
early in their history (i.e. ellipticals).
Secondly, the color of a galaxy is dependent on its star formation
history. The higher the recent star formation rate,
the ``bluer" the galaxy (Bruzual \& Charlot 2003).
Thus, for a given nova rate per unit mass, late-type galaxies
will have a higher $K$-band luminosity-specific nova rate when
compared with early-type galaxies.

In addition to the dependence on Hubble type, Yungelson et al. (1997) pointed
out that galaxy mass was a key factor affecting the LSNR.
In particular, they pointed to work by
Gavazzi \& Scodeggio (1996) who have argued that
giant late-type spiral galaxies ($L_H\sim10^{11}~L_{\odot}$)
have an exponentially decaying star formation
rate ($\tau\sim0.5\times10^9$~yr),
while dwarf late-type galaxies ($L_H\sim10^{8}~L_{\odot}$)
have a more nearly constant star formation rate characterized by
$\tau\grtsim5\times10^9$~yr. Thus
low-mass spirals and irregular galaxies
have had more active star formation in the recent past
compared with the more massive spiral systems.

Overall, the models of Yungelson et al. (1997)
seem to support the conclusions
of Della Valle et al. (1994) who found the highest LSNRs in
the low-mass, late-type galaxies such as the LMC, SMC and M33.
The one galaxy type
missing from the analysis were the massive, late-type spirals, for which no
nova rates had yet been measured.
In an attempt to remedy this deficiency,
Shafter et al. (2000)
undertook nova surveys in the massive late-type
galaxies M51 and M101, with the giant elliptical M87
included for comparison as a massive early-type galaxy.
Ultimately, the Shafter et al. (2000) study and a similar study
targeting M33 by
Williams \& Shafter (2004) failed to confirm any
significant correlation between LSNR and Hubble type.

In recent years, the results of the Shafter et al. (2000)
have been called into question by the work of
Shara and collaborators.
In an analysis of archival HST images of M87,
Shara \& Zurek (2002) reported a (preliminary) nova rate
significantly higher than that found by
Shafter et al. in their ground-based study.
This result prompted Neill \& Shara (2004)
to contemplate the possibility that many ground-based,
multi-epoch nova surveys with sparse temporal coverage
may have systematically underestimated nova rates because
the correction for the effective survey times is often
large and uncertain.

Given the importance of establishing reliable nova rates
for galaxies spanning a wide range of Hubble types, and
for the massive late-type spiral galaxies in particular,
we have undertaken a new survey for novae in M101 using
the 90Prime camera on the Bok 2.3-m reflector.
As a relatively nearby ($\mu_0=29.13\pm0.11$ [Freedman et al. 2001]),
nearly face-on, late-type spiral, M101 provides an ideal candidate
for studying the nova rate and the distribution of nova progenitor
populations.
This survey improves on the one conducted by Shafter et al. (2000)
both by monitoring the galaxy more frequently,
and by improving the corrections for spatial and temporal completeness.
In this paper, we report the results of this new survey.

\section{Observations \& Data Reduction}

Observations of M101 were obtained over
10 epochs from May 2005 to May 2007 using the
Steward Observatory 2.3 m Bok Telescope. All observations were made with 
the 90Prime camera (Williams et al. 2004). The camera consists of
an array of 4 4K $\times$ 4K CCDs mounted at prime focus,
yielding a field of view of $\sim$1 square degree ($\sim$30$\arcmin$ per chip).
The large field of view of the 90Prime camera
allowed us to cover nearly the full disk
of M101 on a single chip.

Following our earlier work (Shafter et al. 2000),
imaging was done through a narrow-band
H$\alpha$ filter with a central bandpass
of 6580 \AA\ and a FWHM of $\sim$80 \AA.
Conducting nova surveys in H$\alpha$ offers
several advantages over broad-band continuum observations.
Shortly after eruption, novae develop strong,
broad ($\gtrsim$1000 km s$^{-1}$) H$\alpha$ emission lines 
that typically require a month or more to decline
by more than 2 mag (Ciardullo et al. 1990a), making observations
in H$\alpha$ ideal for surveys with limited temporal sampling.
H$\alpha$ radiation is also less affected by 
internal extinction than B light,
which is important in dusty spiral galaxies such as M101. 
Thus, H$\alpha$ observations allow novae to be identified
even against a bright galactic background, particularly near the nucleus.

At the distance of M101, a total of $\sim4$ hours of observations
per epoch
were required to detect a significant fraction of the
nova eruptions. Individual exposures of typically 900 seconds were 
taken, typically on a single night, and later median-combined in order
to suppress cosmic ray artifacts and avoid CCD saturation.
Each epoch also
contained a set of bias frames, dark frames, and sky flats
for all 4 CCD chips that were used in the image reduction process. 
After de-biasing and flat-fielding the individual images,
a world coordinate system (WCS) for each
image was assigned using the
MSCTPEAK routine in IRAF\footnote{IRAF (Image Reduction and Analysis
Facility) is
distributed by the
National Optical Astronomy Observatories, which are operated by AURA,
Inc.,
under cooperative agreement with the National Science Foundation}
using coordinates from the U.S. Naval Observatory A2 catalog.
With coordinates accurately assigned, images
from a given epoch were then aligned using several stars common to each
image.
In 3 epochs, M101 was centered on a single chip (chip 1) in all exposures.
This allowed the
images to be aligned using the IMALIGN routine in IRAF. The other 7
epochs followed a procedure of alternating the center of M101 between 3
chips (chip 3 was excluded due
to a large area of bad pixels). In these cases it was necessary to align
the images using SREGISTER due to the slight
amount of rotation from one chip to the next.
After alignment, the images were median-stacked to suppress transient
features.
The resulting master image represented between 2.6 and 4.5 hours of
coverage per epoch. Table~1
contains a summary of the observations.


\subsection{Nova Detections }

To identify novae, two master combined images
from epochs at least one year apart were
aligned and then blinked by eye. Novae were identified by their variability. 
To qualify as a nova,
an object had to appear in at least
one epoch and be absent from previous and/or subsequent epochs (greater
than 3-4 months before or after the date of the image
in which the nova was first seen). This 
method enabled novae to be detected to within 3$'$ of the nucleus and yielded a
total of 6 novae. Due to the brightness of
the background in the nucleus region, it was
necessary to first subtract the background
level before searching for novae in the inner regions. 
An estimate of the background level was obtained with
the MEDIAN routine in IRAF. In this routine, each pixel of the image
was replaced by the median of the pixel values in an $11\times11$
pixel box centered on the pixel in question.
The resulting median image was then subtracted from the original,
allowing novae to be detected to within 10$''$ of the nucleus.
Six additional novae were identified by directly blinking the median
subtracted images. 

As a check on the direct blinking method,
master images were aligned and then the point
spread functions were matched with the PSFMATCH routine.
Each image was divided into 16 $1024\times1024$ pixel
sub-images prior to the alignment. The matched images were
divided by each other and novae were identified as residual sources on the
differenced frame.
This method resulted in the discovery of one new nova,
and allowed us to confirm novae found via directly
blinking the images. A total of 13 novae were discovered
in the 10 epochs spanned by our survey.

\subsection{Nova Photometry}

Photometric measurements were
made of both the novae and secondary standard
stars using two different methods.
In the first case, raw magnitudes were extracted
directly using the PSF routines in the IRAF package DAOPHOT.
As a check on the PSF measurements,
simple-aperture photometry was performed with PHOT. 
Due to the high level of galactic background light around
the novae and the standard stars, we first extracted
$60\times60$ pixel subimages centered around each target.
After excluding a 10 pixel radius aperture around the target,
we fit a two-dimensional surface to the background
with the IMSURFIT routine.
This fitted surface was subtracted from the image,
leaving a flattened
image on which to perform the aperture photometry.
Using apertures of the order of the seeing FWHM,
we derived instrumental magnitudes for the
novae discovered in this survey,
and for the secondary H$\alpha$ standard stars given
Shafter et al. (2000),
which were originally calibrated from the spectrophotometric
standard stars of Stone (1977) and Oke (1974).
The magnitudes resulting from both
the PHOT and PSF methods were usually consistent
to within 0.1 mag and in these cases the
mean of the two measurements was adopted.
When the magnitudes differed by more than 0.1 mag,
the discrepancy was attributed to the difficulty
in fitting the level of the background light and
we adopted the PSF magnitude. 
Following Shafter \& Irby (2001)
and Williams \& Shafter (2004) we have assumed a 100\%
filling fraction over the bandpass of the H$\alpha$ filter.
Although this introduces a small error in the calibration
in cases where a nova's H$\alpha$ emission underfills
the bandpass, it has
the benefit of allowing a direct comparison of our
magnitudes with those of previous studies.

The discovery dates,
positions, and H$\alpha$ magnitudes of
the 13 novae discovered in our survey are presented in Table~2.
Two of the novae, CSM2005-1 and CSM2006-6, were detected
in subsequent epochs. However, since the times of maximum
light for these novae cannot be constrained,
no useful light curve information
can be gleaned from these data.
 
\section{The Nova Rate}

Following Shafter et al. (2000) and Williams \& Shafter (2004),
we have estimated the
nova rate in M101 using two approaches:
through a Monte Carlo simulation, and through
a mean nova lifetime method.
In both cases, a computation of the nova rate requires that we accurately
estimate the completeness of our survey.
Following the procedure of Williams \& Shafter (2004), the
completeness was established with
artificial star tests using
the ADDSTAR routine in IRAF. For a typical survey image, 
ADDSTAR generated artificial novae based on the PSF
model that was generated for the image during
the photometry performed earlier.
Artificial novae for 8 equally-spaced magnitude bins,
from $m_{H\alpha}$ = 19.5 to $m_{H\alpha}$ = 23.0, were
generated. For each magnitude bin, 100
artificial novae were distributed throughout the
image with a spatial distribution that
followed the $K$-band light as determined from
the Two Micron All Sky Survey (2MASS) data (Jarrett et al. 2003).
We then searched for these ``novae'' using the same methods employed
to find the actual novae. This process was repeated
three times, with similar results, and the mean fraction of novae
recovered in each bin allowed us to construct the
completeness function, $C(m)$, shown in Figure~1.
The fraction of novae recovered starts to drop steeply
beyond a magnitude of $m_{H\alpha}=21.5$, where we estimate
the survey is $\sim60$\% complete.

\subsection{The Monte Carlo Method}

In the Monte Carlo method, we compare the actual number of
novae observed in our survey ($N_{obs}=13$) with
estimates of the number of novae we expect to observe
as a function of the intrinsic nova rate, $R$, in M101.
Initially, a set of model 
H$\alpha$ light curves are constructed by selecting
peak magnitudes and decay rates at random from a sample of novae observed in
earlier surveys of M31 (Shafter \& Irby 2001) and M81 (Neill \& Shara 2004).
For a range of plausible values of $R$,
an observed nova luminosity function, $n(m,R)$, is then computed
based on the dates of our survey (see Table~1), and
the distance of M101, which we take
to be $\mu_0=29.13\pm0.11$ (Freedman et al. 2001).
This luminosity function
is then convolved with the completeness function $C(m)$ 
in order to predict the number of novae expected in our survey: 
\begin{equation}
n_{obs}(R) = \int C(m)n(m,R) dm. 
\end{equation}
The Monte Carlo routine is repeated $10^{4}$ times and the number of times
that $n_{obs}(R)$ matches the actual number of novae observed, $N_{obs}$,
is recorded. When normalized, the number of matches as a function of $R$ 
produces a probability distribution function for the nova rate in M101. The peak of
the distribution corresponds to the most probable nova rate,
$R = 11.1_{-1.5}^{+1.9}$ yr$^{-1}$ in the surveyed region
as shown in Figure~2.
The error estimates were determined from the probability distribution
where the computed nova rate lies between 9.6 and 13.0 novae per year
with 50\% probability.

\subsection{The Mean Nova Lifetime Method}

As a check on the Monte Carlo results, a rough estimate of the 
nova rate can also be obtained using the mean nova lifetime method
(Ciardullo et al. 1990a),
which is based on the method used by Zwicky (1942) to determine extragalactic 
supernova rates.
The nova rate in the surveyed region of the galaxy can be expressed as
\begin{equation}
 R = \frac{N(M < M_c)}{T(M < M_c)},
\end{equation}
where $N(M < M_c)$
is the number of novae observed brighter than a specified cut-off magnitude
$M_c$, and $T(M < M_c)$ is a quantity
known as the ``effective survey time".
For a multi-epoch survey,
the effective survey time depends on the sampling frequency
of the survey and on the mean nova lifetime, $\tau_c$, which
is length of time a typical nova remains brighter than $M_c$.
We have
\begin{equation}
 T(M < M_c) = \tau_c + \sum_{i=2}^{n} min(t_{i} - t_{i-1}, \tau_c),
\end{equation}
where $t_i$ is the time of the $i$th observation.
A simple relationship between $\tau_c$ and $M_c$ was calibrated by
Shafter et al. (2000)
based on observations of novae in the bulge of M31, and is adopted here:
\begin{equation}
\rm log\ \tau_c(days) \simeq 6.1(\pm0.4) + 0.56(\pm0.05)M_c.
\end{equation}

Before we can compute a nova rate, we must specify a suitable
cut-off magnitude $M_c$ for our survey.
As mentioned earlier, the survey completeness drops off precipitously
beyond an apparent magnitude of $m_{H\alpha}=21.5$. Thus, for
our purposes we will adopt a value of $M_c$ corresponding to $m_c=21.5$.
Using a distance modulus of $\mu_{o}$ = 29.13$\pm$ 0.11 for M101
(Freedman et al. 2001), and estimated foreground extinction of
$\sim0.05$~mag at H$\alpha$ (Schlegel et al. 1998), we
find $M_c=-7.7\pm0.24$. Equations~3 and~4 (coupled with the survey times
listed in Table~1) then yield $\tau_{c}=61.3\pm3.7$~days, and
$T(M < M_{c})$ = $479\pm15$ days for the mean nova lifetime
and the effective survey time, respectively.
Given that
eight of the 13 novae detected were brighter than our
adopted cut-off magnitude of $m_c=21.5$, where we are $\sim60$\%
complete, equation~2 yields a nova rate of $10.2\pm4.0$. Despite
the simplicity of the mean nova lifetime approach, which does not
properly take the incompleteness of the survey into account through
$C(m)$, this result is in surprisingly good
agreement with the results of the Monte Carlo calculation.

\subsection{The Global Nova Rate}

The Monte Carlo and mean nova lifetime nova rates reflect the nova rates
in the surveyed region of M101, and must be corrected for any fraction
of M101 that falls outside the coverage of our survey. Fortunately,
the large field of view provided by the 90Prime camera has enabled
us to cover essentially all of M101.
Using $K$-band photometry derived
from 2MASS data (Jarrett et al. 2003), we estimate
that our effective survey area
includes $\sim$95\% of the total infrared luminosity of M101.
Thus, our estimate of the global nova rate in M101 based on the
Monte Carlo calculation is $11.7_{-1.5}^{+1.9}$~yr$^{-1}$, which,
given the uncertainties inherent in our calculations, is in remarkable
agreement with the results of Shafter et al. (2000) who estimated
a rate of $12\pm4$ yr$^{-1}$ based on their more limited survey.
We note that our nova rate estimates should more properly be
considered as lower limits on the true nova rate in M101 since
we have not made corrections for the effects of extinction
internal to M101, which varies significantly with spatial position.

\subsection{The Luminosity-Specific Nova Rate}

In order to compare the nova rates in different galaxies,
the luminosity-specific
nova rate (LSNR) is used to approximate a normalization by stellar mass. 
The infrared luminosity is used as a proxy for mass because
it is a better tracer
of low-mass stars than visible light.
Thus, extragalactic nova studies typically normalize 
the global nova rate to the total $K$-band luminosity
of the galaxy (Ferrarese et al. 2003; Williams \& Shafter 2004).
We parameterize our LSNR, $\nu_{K}$, as the number of novae
per year per $10^{10}~L_{\odot}$ in the $K$-band. 

The LSNR was determined using two different estimates for the
$K$-band luminosity: from the integrated $B$ magnitude
and $B-K$ color (as in Shafter et al. 2000),
and from the value determined by the 2MASS Large
Galaxy Atlas (Jarrett et al. 2003).
Though the 2MASS data provides directly measured $K$ magnitudes
for all galaxies with determined nova rates,
there are systematic discrepancies between the 
$K$ magnitudes from 2MASS (K$_{2MASS}$) and those from $B-K$ colors
(K$_{color}$) (Ferrarese et al. 2003; Williams \& Shafter 2004).
Williams \& Shafter attribute the discrepancy to difficulties
in measuring the background levels in galaxies
that cover a large area on the sky.
Following these authors, we consider the LSNR derived
from galaxy colors to be more reliable than the LSNR from 2MASS.

For our nova rate of $11.7^{+1.9}_{-1.5}$ yr$^{-1}$
and the $K_{color}$ magnitude of 5.02 $\pm$ 0.14 
we find
$\nu_{K,color}=1.23\pm0.27$ novae per year per $10^{10}~L_{\odot,K}$,
while the value of $K_{2MASS}=5.51\pm0.05$ yields
$\nu_{K,2MASS}=1.94\pm0.42$ novae per year per $10^{10}~L_{\odot,K}$.
Both of these estimates are consistent with the
value of $\nu_{K,color}=1.27\pm0.46$
novae per year per $10^{10}~L_{\odot,K}$
given in Williams \& Shafter (2004), and based on the
Shafter et al. (2000) study.

\section{The Spatial Distribution of M101 Novae}

It is possible that the spatial distribution of novae within
a galaxy may yield information about the evolution of binary systems
in that galaxy. For example, some studies of novae in elliptical
galaxies have found that novae are more heavily concentrated in the
central regions (e.g., Madrid et al. 2007).
The spatial distribution
of the 13 novae discovered in the present study are plotted
over mean
$K$-band isophotes of M101 derived from 2MASS data in Figure~3.
Each isophote represents a 10\% 
change in the total light of the galaxy.
Novae brighter than the adopted limiting magnitude are
represented by closed circles and those that were fainter
than the limit are open circles. The novae found in the M101
study of Shafter et al. (2000) are also included and are
represented as squares.
It is interesting to note that
although the previous study had a field-of-view of
only $\sim16'\times\sim16'$ (smaller than the estimated
$\sim27'$ angular diameter of M101),
none of the novae detected in the present survey with its
larger coverage ($30'\times30'$) fell outside the limits of our
earlier survey.

Figure~4 shows the cumulative distribution of all novae from the
current study and from Shafter et al. (2000) that were
brighter than the limiting magnitude of their respective
surveys compared to the
normalized background $K$ light of the galaxy. A Kolmogorov-Smirnoff (KS)
test indicates that the nova distribution does not differ
significantly from that of the background light (KS=0.94).
This result, as well as similar results for other spiral
systems such as M31 (Shafter \& Irby 2001), M33 (Williams \& Shafter 2004),
and M81 (Neill \& Shara 2004), differs from the results found in M87
(Madrid et al. 2007) and M49 (Ferrarese et al. 2003) where the novae
appear concentrated towards the centers of the galaxies,
and indicates that nova progenitors
may be more uniformly distributed in spiral galaxies than in ellipticals. 

\section{Discussion}

\subsection{The Nova Rate in M101}

In the Shafter et al. (2000) study a nova rate of $12\pm4$~yr$^{-1}$,
corresponding to a LSNR of
$\nu_K=1.27\pm0.46$ novae per year per $10^{10}~L_{\odot,K}$
was determined based on one epoch of observation in each of four
consecutive years between 1994 and 1997. The nova rate was computed
using the mean nova lifetime method, and a simplified Monte Carlo
routine. Both assumed a sharp cut-off in nova detectability at a
limiting magnitude of $m_{H\alpha}=22.0$.
Given that the frequency of observation was just one epoch per year,
the effective survey time, $T(M<M_c),$ was determined solely by the
(inherently uncertain) value of the mean nova lifetime, $\tau_c$,
causing systematic errors in the mean nova lifetime relation
$\tau_c(M_c)$ to propagate directly into the computed nova rate.
Errors in the mean nova lifetime relation arise from the fact that
it was derived from a limited sample of M31 novae with
measured H$\alpha$ light curves from the studies
of Ciardullo et al. (1990a) and Shafter \& Irby (2001).
Without including well-sampled H$\alpha$
light curves from late-type galaxies
we cannot state with certainty that
this relationship holds for M101 novae.
The Monte Carlo nova rate estimate was also affected by these uncertainties
as this method also used light curve data from M31 novae.

The present study has attempted to
improve on the earlier survey in a number of ways.
We began by increasing the sampling frequency
to include multiple epochs per year, in an attempt to minimize the
dependence of the final nova rate on uncertainties in the
mean nova lifetime relation. In the present survey
we obtained ten epochs in a three year period, compared with the
previous study, which obtained only one epoch per year for four years.
Secondly, instead of adopting a
specific limiting magnitude for the Monte Carlo nova rate calculation,
we estimated the nova completeness as a function of magnitude, $C(m)$,
using artificial star simulations. Finally, we further improved
the Monte Carlo simulation by including H$\alpha$ light curves
of novae in M81 that have become available
from the study of Neill \& Shara (2004).
Despite all of these improvements, and
despite the uncertainties inherent in the use of multi-epoch surveys,
a comparison of our results with those of Shafter et al. (2000)
indicates that the derived nova rate for M101 (11.7~yr$^{-1}$)
is a robust one, in good agreement
with the rate of 12~yr$^{-1}$ found in the previous study.

Despite the agreement between the two surveys,
a nagging concern is that our temporal coverage, while improved
over the original Shafter et al. (2000) survey, may still
result in an underestimate of the nova rate. The only H$\alpha$
survey with nearly continuous temporal coverage over an
extended period of time has been
the $\sim5$ month long survey of M81 reported in Neill \& Shara (2004).
This survey resulted in a LSNR nearly double that
reported a decade earlier by Moses \& Shafter (1993),
exacerbating the concern that surveys with sporadic temporal coverage
may be systematically underestimating nova rates. It is worth pointing
out, however, that much of the boost in the LSNR was not the result of an
increase in the nova rate, per se, but because Neill \& Shara (2004) adopted
a different (and lower) $K$ band luminosity for M81 based on
the 2MASS survey (Jarrett et al. 2003).
As discussed in Williams \& Shafter (2004) there is evidence that the
2MASS survey may have systematically
underestimated the integrated $K$-band luminosities for nearby
galaxies with relatively large angular diameters, such as M81. 
If we compare the nova rates from the two surveys, we find that
Neill \& Shara's value of $33^{+13}_{-8}$~yr$^{-1}$ is only
$\sim40$\% higher than a preliminary estimate of $24\pm8$~yr$^{-1}$
reported by Moses \& Shafter (1993). Thus, although
surveys with continuous coverage may result in a modest increase in the
nova rate for some galaxies, they are unlikely to
fundamentally alter the conclusion that the LSNRs are not
strongly dependent on Hubble type.

Perhaps a more significant concern
is whether the properties
of the H$\alpha$ light curves from M31 and M81 used in the
Monte Carlo simulation are representative
of those in other galaxies such as M101
with differing Hubble types. In this regard,
it would still be very useful to undertake continuous
H$\alpha$ surveys not only in M101, but in ellipticals
such as M87 as well, in order to obtain H$\alpha$ light curves
from novae arising from different stellar populations.

\subsection{The LSNR and Hubble Type: Latest Thoughts}

Williams and Shafter (2004) recently summarized the available
data for extragalactic nova rates, and
found little support for the notion
that the LSNR correlates strongly with Hubble type
(see their figure~6).
The results for M101 presented here,
which are in excellent agreement with
the nova rate for this galaxy found previously,
obviously do not alter this conclusion. 
While it
appears likely that the Large and Small Magellanic clouds
have LSNRs $2-3$ times larger than other galaxies, in agreement
with the predictions of Yungelson et al. (1997), the LSNRs
in the Virgo ellipticals remain highly uncertain.

Arguably, the most uncertain nova rate for an individual
galaxy concerns the nova rate in M87.
Shafter et al. (2000)
estimated a rate of $91\pm34$ based on ground-based observations
that were blind to novae within a radius of $\sim$25$''$ of the nucleus.
To determine a global nova rate for the galaxy, these authors
assumed that the cumulative nova distribution
followed the integrated light of the galaxy and then extrapolated
from the novae found in the galaxy's outer regions. Recent
HST observations by Madrid et al. (2007)
strongly suggests that this assumption may not be valid.
These authors found a total of 13 nova candidates
during a HST STIS survey that spanned three epochs over 51 days. The
survey covered a $\sim24.7''\times24.7''$ region centered
on the nucleus of M87. The novae were highly concentrated
toward the center of the field, with 11 of the 13 novae lying within
$9''$ of the nucleus. After correcting for the survey time,
they found a nova rate in the surveyed region of $\sim$64~yr$^{-1}$.
This rate is amazingly high, considering that the surveyed region
covered only $\sim$6\% of the galaxy's $K$-band luminosity.
Extrapolation to the entire galaxy would suggest a nova rate
of more than 1000~yr$^{-1}$! Given that it is extremely unlikely
that the ground-based surveys could have missed such a large
nova population, it seems likely that the nova distribution
in M87 is significantly more centrally condensed
compared with the integrated background light.
If so, the M87 nova rate estimated
from the ground by Shafter et al. (2000) would clearly
underestimate the true nova rate.

Further evidence for a high M87 nova rate can be found in the
preliminary results reported
by Shara \& Zurek (2002). These authors analyzed 10
epochs of archival WFPC2 images
and claim to have detected over 400 novae within
a radius of $\sim90''$ of M87's nucleus.
Perhaps surprisingly, unlike
Madrid et al. (2007), Shara \& Zurek found that the spatial
distribution of their
nova candidates followed the background light reasonably well.
Although at the time of publication,
these authors had not yet fully assessed the completeness of their
survey, they argued for a lower limit on the global
nova rate of 300~yr$^{-1}$.

The only other Virgo elliptical for which a nova rate is available
based on HST observations is M49 (Ferrarese et al. 2003).
Despite the fact that M49 is slightly more luminous
than M87 (M49 is the brightest galaxy in the Virgo cluster),
Ferrarese et al. determined a nova rate of only $\sim$100~yr$^{-1}$.
Although this discrepancy seems hard to understand,
given that both galaxies have similar luminosities and Hubble types,
there is one significant way in which these two galaxies differ:
the specific frequency of globular clusters in M87 is $\sim$3 times
higher than that in M49 (Kissler-Patig 1997, Rhode \& Zepf 2001).
The presently uncertain ratio of nova rates and its similarity to the
ratio of globular cluster frequencies between these
galaxies raises the interesting possibility that
a significant fraction of the nova
binaries in elliptical galaxies may be spawned in globular clusters, as was
originally proposed by Ciardullo et al. (1987)
to explain the high bulge rate in M31.

\section{Conclusions}

An H$\alpha$ survey for novae in M101 was carried out in 10 epochs
over a period of $\sim$3~yr, with a total of 13 novae identified.
After correcting for the effective survey time, we arrive
at a global nova rate of $11.7^{+1.9}_{-1.5}$~yr$^{-1}$, which
corresponds to a luminosity-specific nova rate of
$1.23\pm0.27$ novae per year per $10^{10}~L_{\odot,K}$ when
the $K$ luminosity is
derived from B$_{tot}$ and the $B-K$ color,
and $1.94\pm0.42$ novae per year per $10^{10}~L_{\odot,K}$
when the $K$ magnitude is taken from the 2MASS measurements
(Jarrett et al. 2003).
Both values are consistent, within the stated uncertainties,
with the previously published value of
$1.27\pm0.46$ novae per year per $10^{10}~L_{\odot,K}$
given in Williams \& Shafter (2004). As in previous
determinations of the nova rates in spiral galaxies, a
major source of uncertainty in the derived rates is introduced
by our inability to properly account for the effects of extinction
internal to M101. In this regard our estimates for the global nova
rate and the LSNR in M101 could more conservatively
be considered as lower limits to their true values.

Future observations should be focused on
addressing three remaining sources of uncertainty in computing
the LSNR. As described earlier, more frequent temporal sampling
would be useful, not only to minimize uncertainty in computing
the effective survey time, but also in establishing a larger
sample of template light curves to be used in the
Monte Carlo simulations. The availability of light curves from
novae arising from galaxies with a wide range of Hubble types will establish
whether the properties of the light curves are sensitive to
stellar population. In this regard, we expect that synoptic
survey telescopes such as the PanStarrs and
the LSST will revolutionize the study of novae in extragalactic
systems. Finally, there remains the possibility that a
significant fraction of the novae, particularly in late-type
galaxies such as M101, may be missed because of extinction
within the galaxies themselves. Perhaps future observations with
wide-field infrared imagers can help alleviate this concern.

\acknowledgments

We acknowledge support through NSF grant AST-0607682 to AWS.

\clearpage
\begin{deluxetable}{lcccc}  
\tablecaption{Summary of Observations}
\tablewidth{0pt}
\tablehead{	\colhead{} 	& 
		\colhead{Julian Date}		& 
		\colhead{Number of}		& 
		\colhead{Total Integration Time}		& 
		\colhead{Mean Seeing} 		\\
		\colhead{UT Date} 		& 
		\colhead{(2,450,000+)} 		& 
		\colhead{Exposures}		&
		\colhead{(hours)} 		& 
		\colhead{(arcsec)} }
\startdata
 2005 May 02\dots & 3493 & 7  & 1.75 & 1.5\\ 
  2005 May 03\dots & 3494 & 9  & 2.25 & 2.0 \\ 
  2005 May 31\dots & 3522 & 4  & 1.33 & 1.4\\
  2005 June 01\dots & 3523 & 8  & 2.67 & 1.5\\
  2006 Feb 07\dots & 3774 & 11  & 2.75  & 2.0\\
  2006 Feb 21\dots & 3788 & 7 & 1.75 & 1.3 \\
  2006 Feb 22\dots & 3789 & 5  & 1.25 & 2.0 \\
  2006 Apr 19\dots & 3845 & 12  & 3.0 & 1.4\\ 
  2006 Apr 20\dots & 3846 & 4 & 1.0 & 1.4\\ 
  2006 May 24\dots & 3880 & 18  & 4.5 &1.8\\
  2006 Jun 17\dots & 3904 & 10  & 2.5  & 1.2\\
  2006 Jun 18\dots & 3905 & 4 & 1.0  & 1.2\\ 
  2006 Dec 13\dots & 4083 & 4  & 0.67  & 1.5\\ 
  2007 Jan 28\dots & 4129 & 13  & 3.25  & 1.2\\ 
   2007 Mar 11\dots & 4171 & 14  & 3.5  &2.5\\ 
  2007 Mar 12\dots & 4172 & 4 & 1.0  & 2.0\\
  2007 May 22\dots & 4244 & 9  & 2.25 & 1.5\\
  2007 May 23\dots & 4245 & 9  & 2.25 & 1.0\\
\enddata
\tablecomments{All exposures were taken through an H$\alpha$ filter.}   
\label{tab1}
\end{deluxetable}


\begin{deluxetable}{lccccc}  
\tablecaption{M101 Nova Positions and Magnitudes}
\tablewidth{0pt}
\tablehead{	\colhead{} 	& 
		\colhead{JD Discovery}      &
		\colhead{$\alpha$}		& 
		\colhead{$\delta$}		& 
		\colhead{$\Delta r $}		& 
		\colhead{$m_{H\alpha}$} 		\\
		\colhead{Nova} 		& 
		\colhead{(2,450,000+)} 		& 
		\colhead{(J2000.0)} 		& 
		\colhead{(J2000.0)} 		& 
		\colhead{(arcmin)} 		& 
		\colhead{(mag)} }
\startdata
 CSM2005 - 1\dots & 3493  & 14 03 27 & 54 26 32 & 6.04   & 20.7 \\ 
                  & 3522  & \dots    & \dots    & \dots  & 21.6 \\ 
 CSM2005 - 2\dots & 3522  & 14 03 27 & 54 23 03 & 3.03 & 21.5 \\ 
 CSM2006 - 1\dots & 3788  & 14 03 02 & 54 21 23 & 1.60 & 21.8\\ 
 CSM2006 - 2\dots & 3845  & 14 03 20 & 54 20 27 & 1.17 & 22.1 \\ 
 CSM2006 - 3\dots & 3880  & 14 03 24 & 54 18 39 & 2.78 & 21.0 \\  
 CSM2006 - 4\dots & 3880  & 14 03 59 & 54 26 49 & 8.98 & 22.7 \\
 CSM2006 - 5\dots & 3904  & 14 02 40 & 54 15 37 & 7.08  & 21.5\\
 CSM2006 - 6\dots & 4083  & 14 02 53 & 54 23 36 & 3.92  & 19.8 \\
                  & 4129  & \dots    & \dots    & \dots & 21.4 \\
 CSM2007 - 1\dots & 4129  & 14 03 38 & 54 16 30 & 5.73  & 22.0\\
 CSM2007 - 2\dots & 4171  & 14 03 05 & 54 18 57 & 2.22 & 19.8\\
 CSM2007 - 3\dots & 4171  & 14 03 12 & 54 20 43 & 0.17  & 20.2 \\
 CSM2007 - 4\dots & 4171  & 14 03 19 & 54 19 56 & 1.33 & 19.9\\
 CSM2007 - 5\dots & 4244  & 14 03 05 & 54 24 27 & 3.70  & 22.6   \\ 
\enddata
\tablecomments{Units of right ascension are hours, minutes, and seconds, and units of declination 
are degrees, arcminutes, and arcseconds. $\Delta r$ is the distance of the nova
from the center of M101.
We estimate the positions to be accurate to $\sim$1~arcsec
in each coordinate.}
\label{tab2}
\end{deluxetable}
\clearpage

\begin{figure}
\epsscale{1.0}
\plotone{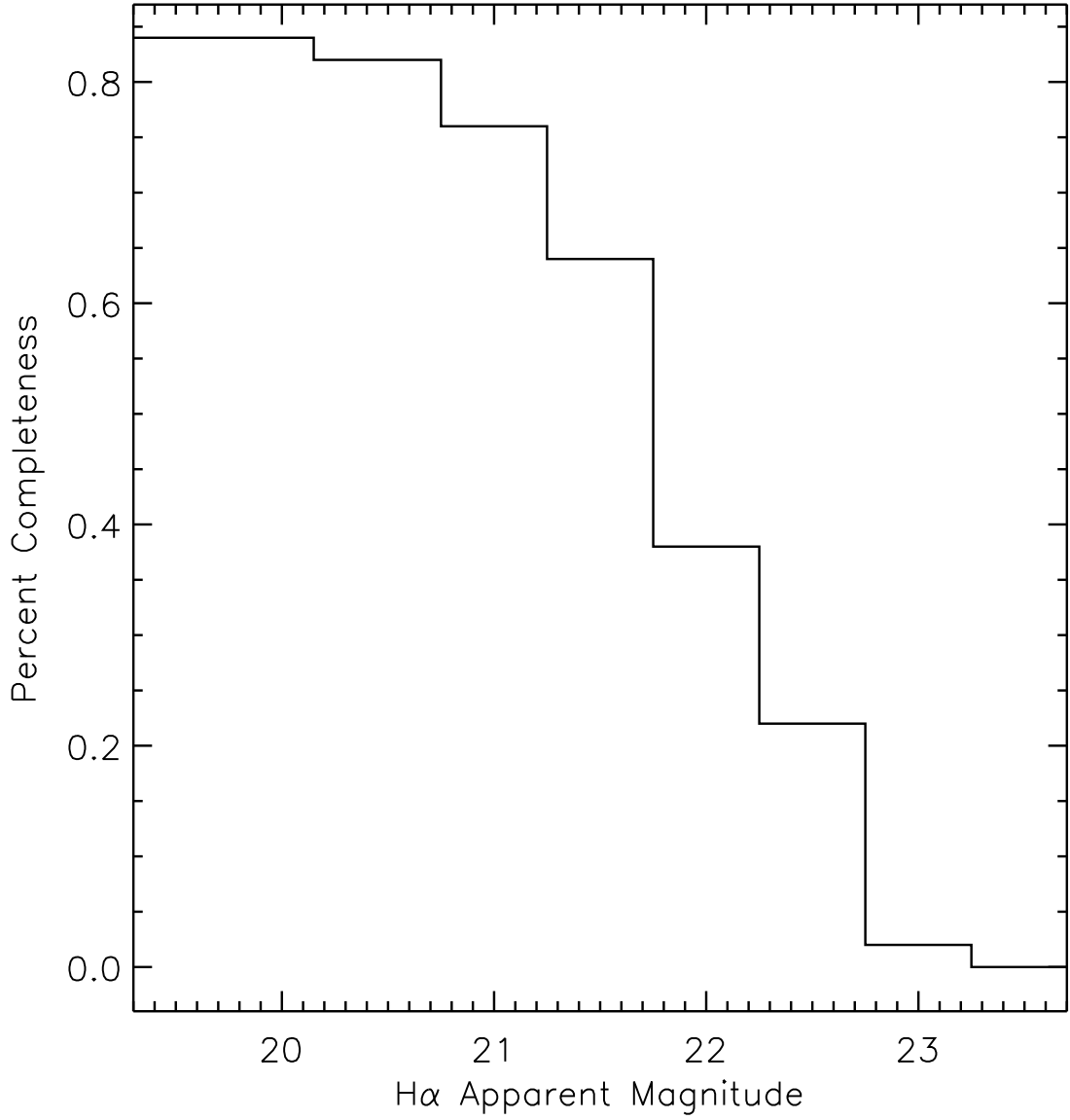}
\caption{Fraction of artificial novae recovered during artificial
star tests as a function of magnitude, $C(m)$.
The cut-off magnitude of $m_{H\alpha}$=21.5 used in the mean
nova lifetime nova rate calculation is
based on the sharp drop in completeness at fainter magnitudes.}
\end{figure}

\begin{figure}
\epsscale{1.0}
\plotone{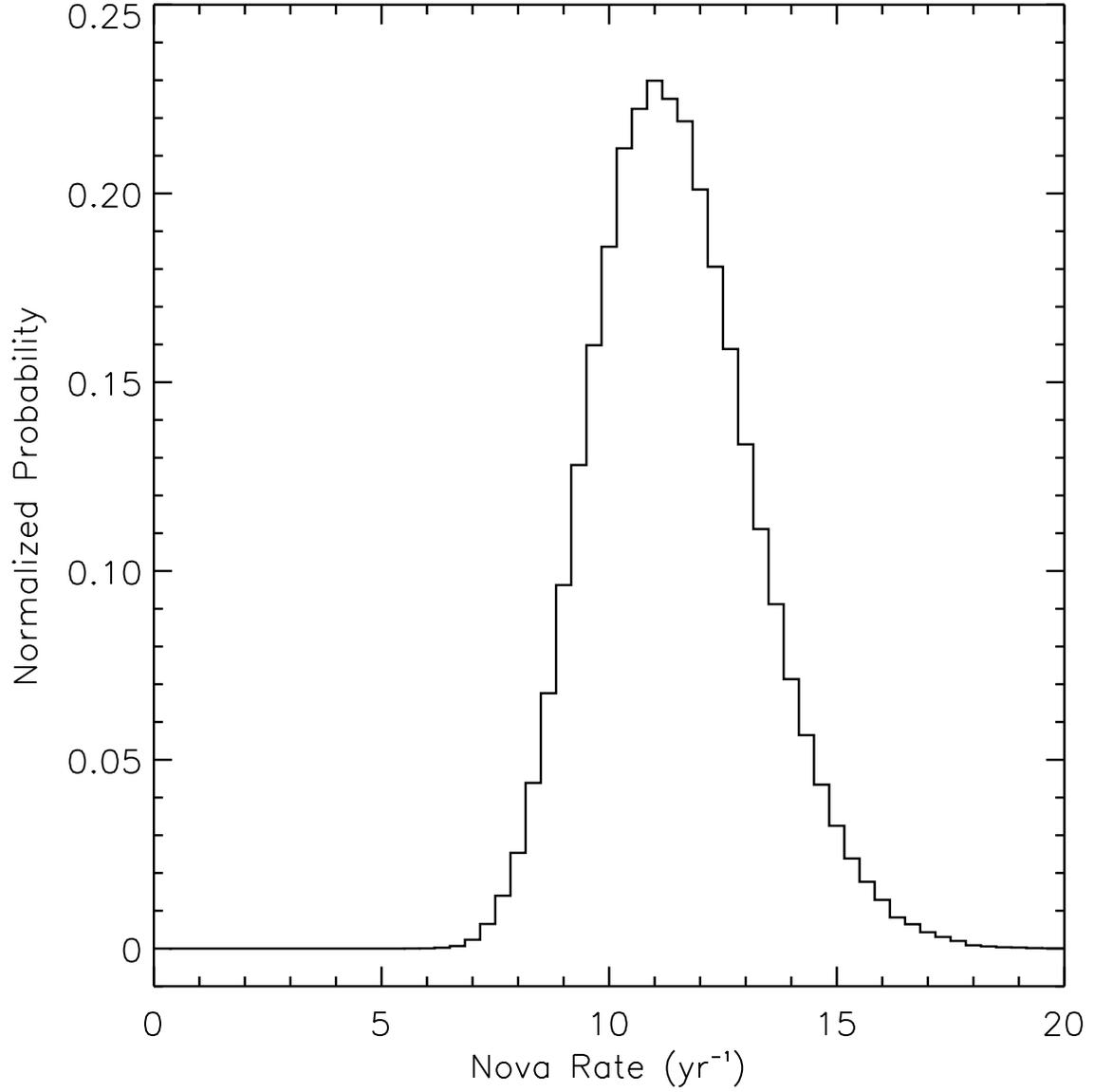}   
\caption{Results of the Monte Carlo simulation.
The peak in the normalized probability distribution at $\sim11$~yr$^{-1}$
represents the most likely nova rate in M101.}
\end{figure}

\begin{figure}
\epsscale{1.0}
\plotone{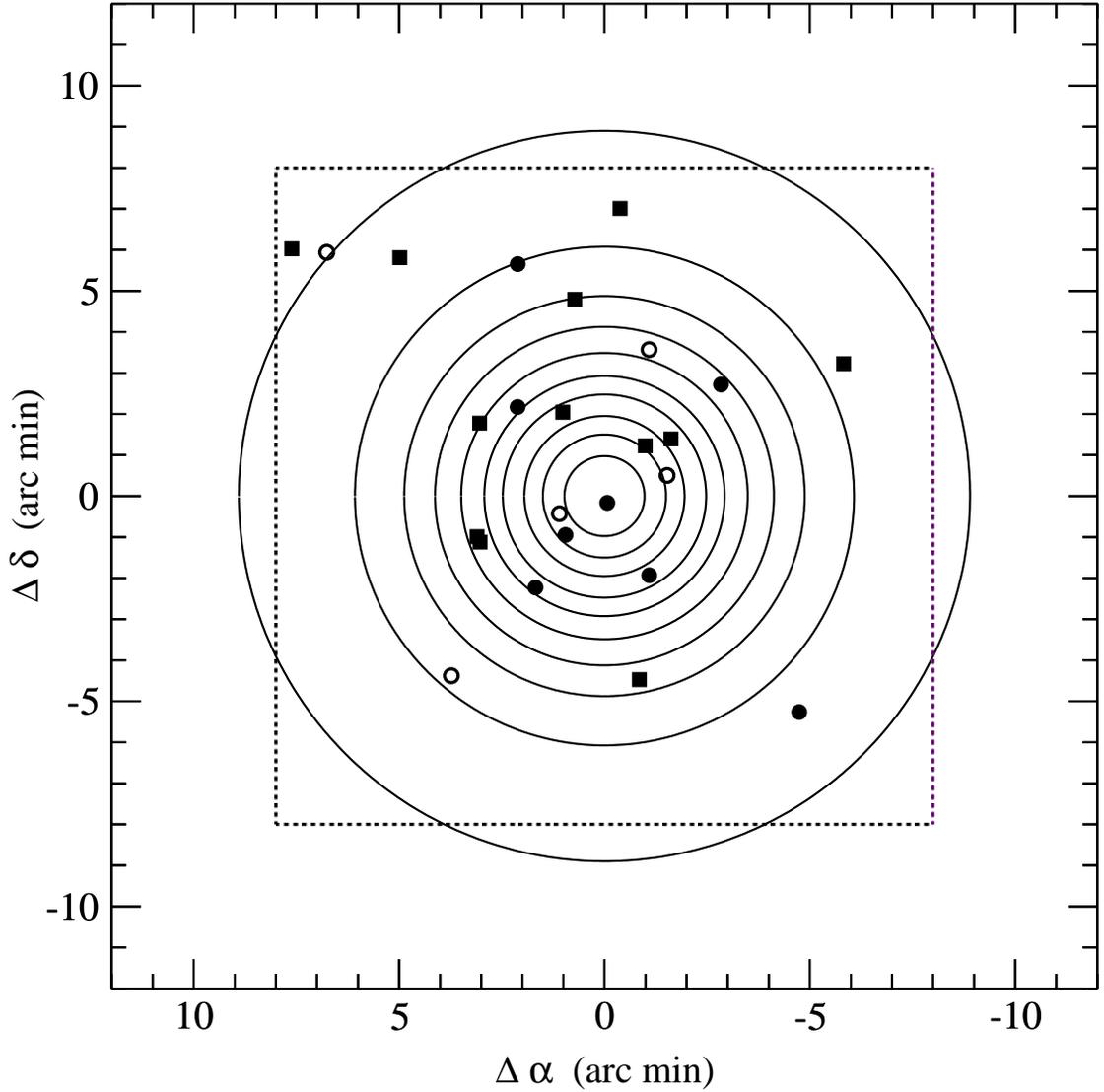}
\caption{The spatial distribution of novae in M101 plotted over
mean $K$-band isophotes derived from 2MASS (Jarrett et al. 2003).
Filled circles represent novae found in the current study, with
open circles indicating
novae that were below the adopted cut-off magnitude,
$m_{H\alpha}=21.5$.
Squares represent novae found in the previous study by Shafter et al. (2000),
with the large broken square delineating the area covered
in that earlier survey.}
\end{figure}

\begin{figure}
\epsscale{1.0}
\plotone{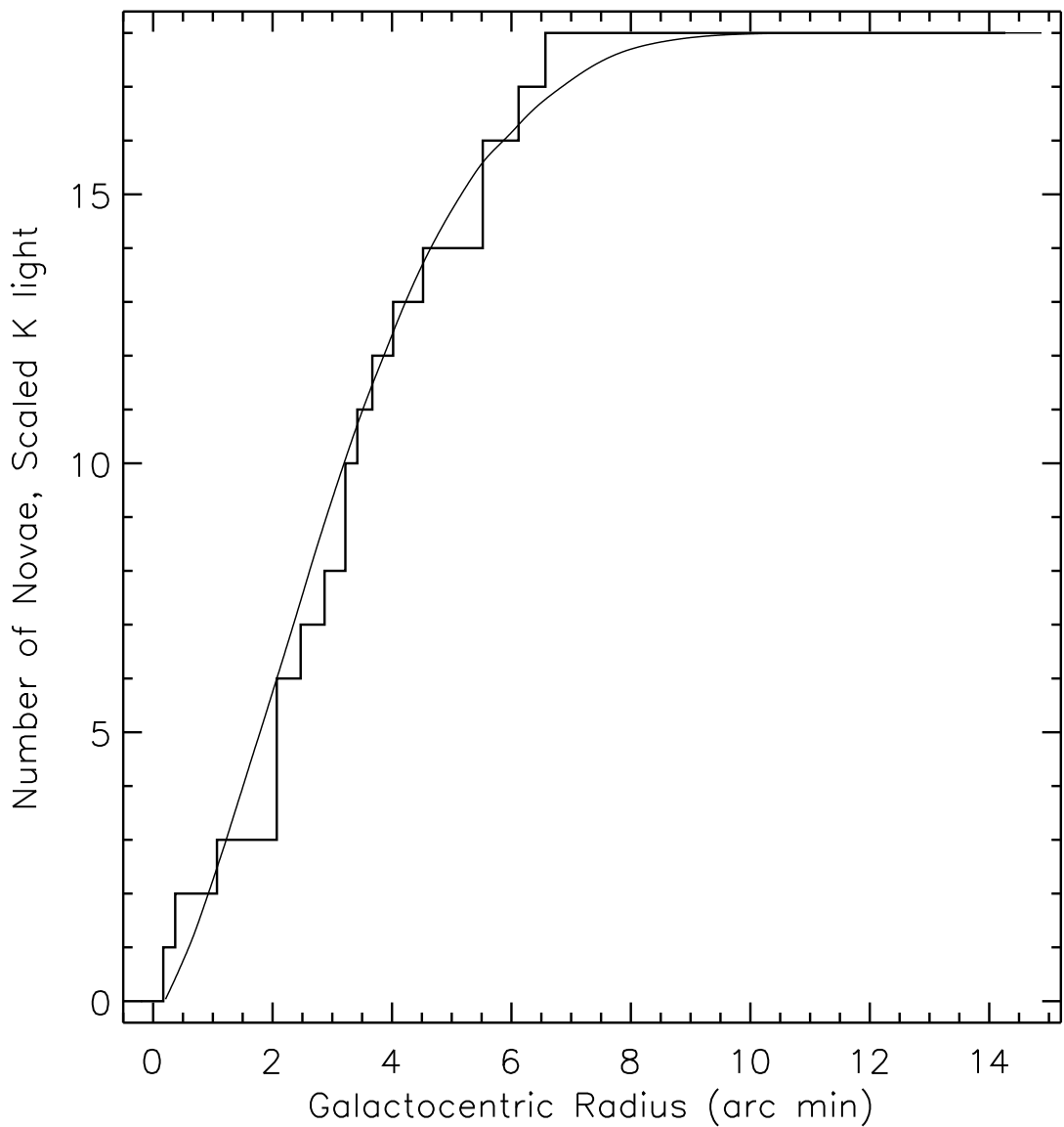}
\caption{The cumulative distribution of the 8 novae
detected above the limiting magnitude along with the
10 novae from Shafter et al. (2000) compared to the 
total distribution of $K$-band light (smooth curve).
Although the number of novae detected is small, the radial distribution
of novae follows the galactic light very well (KS = 0.94).}
\end{figure}

\end{document}